\documentclass[onecollarge,natbib]{svjour2}
\bibpunct{[}{]}{;}{n}{}{,} 
\smartqed  
\usepackage{graphicx}
\usepackage{amsmath,amssymb}
\journalname{Few-Body Systems}
\begin{document}

\title{Heavy meson decay in three-mesons and FSI
}

\titlerunning{Heavy meson decay in three-mesons and FSI}        

\author{T. Frederico, K. S. F. F. Guimar\~aes, O. Louren\c{c}o,  W. de Paula, I. Bediaga, and A. C. dos Reis}


\institute{T. Frederico and W. de Paula\at
Departamento de F\'isica, Instituto Tecnol\'ogico de Aeron\'autica, 12228-900, S\~ao
Jos\'e dos Campos, SP, Brazil
\and
K. S. F. F. Guimar\~aes \at Instituto de Astronomia, Geof\'\i sica e Ci\^encias
Atmosf\'ericas, 05508-900, S\~ao Paulo, SP, Brazil
\and
O. Louren\c{c}o\at
Departamento de Ci\^encias da Natureza, Matem\'atica e Educa\c
c\~ao, CCA, Universidade Federal de S\~ao Carlos, 13600-970, Araras, SP, Brazil
\and
I. Bediaga and A. C. dos Reis \at
Centro Brasileiro de Pesquisas F\'isicas, Rio de Janeiro, 22290-180, RJ, Brazil
}

\date{}

\maketitle

\begin{abstract}
The final state interaction (FSI) contribution to charged $D$ decay into $K\pi\pi$ is
computed within a light-front framework, considering $S$-wave $K\pi$ interactions in $1/2$
and $3/2$ isospin states. The convergence of the rescattering series is checked computing
terms up to the third perturbative order. The role of the resonances above $K^*_0(1430)$,
and the contribution of the $K\pi$ $3/2$ isospin channel to charged three-body $D$
decays, are studied against the available phase-shift analysis.  
\end{abstract}

\section{Introduction}
\label{intro}

The phases of the $K\pi$ elastic amplitude in partial waves from $\ell=$ 0, 1 and 2 were
known from experiments using the reactions  $K^\pm p \to K^\pm \pi^+ n$ and $K^\pm
p\to K^\pm\pi^-\Delta^{++}$ at $13$~GeV~\cite{EstNPB78} and the reaction $K^-p\to
K^-\pi^+n$ at $11$~GeV/c~\cite{AstNPB88}. However, these reactions do not access the low
mass region close to the threshold and  information on the $K\pi$ phase-shifts become
available after the analysis of the charged  $D\rightarrow K\pi\pi$  decays from from E791
\cite{AitPRL02,E791} and FOCUS \cite{FOCUS2} collaborations. These phase-shift analyses
can provide information on the $K\pi$ scattering amplitude starting at the $K\pi$
threshold covering the allowed phase-space for the charged $D$ decay, once three-body
rescattering contributions are under control. In this contribution, we further explore
theoretically the $S$-wave interaction in the different $K\pi$ isospin states in the
calculation of the final state  interaction in  $D^{\pm}\rightarrow
K^{\mp}\pi^{\pm}\pi^{\pm}$ decay channel within a light-front relativistic three-body
model.  

Our work follows  the relativistic three-body model for the
final state interaction in the charged  $D\rightarrow K\pi\pi$ decay based on the
three-meson Bethe-Salpeter equation \cite{karinnpb,MagPRD11,MagPOS12}. In the model developed
here, the decay amplitude is separated into a smooth term and a three-body fully
interacting contribution, which is factorized in the standard two-meson resonant amplitude
times a reduced complex amplitude for the bachelor meson, that carries the effect of the
three-body rescattering mechanism. The off-shell bachelor reduced amplitude is a solution
of an inhomogeneous Faddeev type integral equation, that has as input the $S$-wave isospin
$1/2$  and $3/2$ $K^{\mp}\pi^{\pm}$ transition matrix. In the previous work
\cite{MagPRD11}, the three-body
rescattering calculation took into account  the isospin $1/2$ $S$-wave $K\pi$ interaction
within a chiral model fitted to the LASS phase-shift analysis~\cite{AstNPB88} up to the
$K^*_0(1430)$ resonance, while the available phase-space for the $D$ decay covers energies
up to $1.89$~GeV. Our work extends the previous three-body rescattering model for charged
$D$ decays in order to include the $S$-wave two-body $K\pi$ amplitude in both isospin
states, $1/2$ and $3/2$, for $K\pi$ masses up to $1.9$~GeV. We study the role of the
resonances above $K^*_0(1430)$ in addition to the contribution of the $K\pi$ isospin $3/2$
interaction against the available phase-shift analysis of the charged $D\rightarrow
K\pi\pi$ decay from E791 \cite{AitPRL02,E791} and FOCUS \cite{FOCUS2}  collaborations.

\section{ Decay Amplitude for $D^\pm\to K^\mp\pi^\pm\pi^\pm$}
\label{sec:dalitzd}

The standard technique to study the resonant structure of the three-body decay is the
Dalitz plot analysis. In such a plot, the final state of a particle $P$ decaying into
three particles ($d_{1},d_{2},d_{3}$) can be described through a 
bi-dimensional diagram, where the axis are the two-body invariant masses squared.
The density of events in this plot is given by
\begin{eqnarray}
 d\Gamma(P \to d_{1}d_{2}d_{3}) \propto \frac{1}{M^{3}_{P}}\, |{\cal A}|^{2} \, ds_{12}\, ds_{13}
\end{eqnarray}
and, in such a diagram, we have access to the matrix element ${\cal A}$. The phase-space density of this 
three-body decay, ${M^{-3}_{P}}$, is constant and if we write $M_P$ in terms of the Mandelstam invariants $s_{12}$ and $s_{13}$,
 the kinematically allowed event region is delimited. The structure observed in 
the $D^{+}\rightarrow K^{-}\pi^{+}\pi^{+}$ Dalitz plot \cite{E791}   is an outcome of the decay dynamics and resonances in the decay channels.

The decay amplitude is expressed as a sum over the partial-wave  contributions
\begin{eqnarray}
{\cal A} =\sum_L\left( a_{L}(s_{K\pi})e^{i\phi_{L}(s_{K\pi})}P_L +
a_{L}(s_{K\pi^\prime})e^{i\phi_{L}(s_{K\pi^\prime})} P_L  \right)
\end{eqnarray}
with amplitude $ a_{L}(s_{K\pi})$, phase $\phi_{L}(s_{K\pi})$. In the above sum $P_L$ is a short-hand notation for the angular functions. 
We will be interested in the $S$-wave amplitude $A_{0}(s_{A},s_{B})$.

\section{The $S$-wave $K\pi$  Amplitude}
\label{sec:kpiampl}

In our model the input for the calculation of the $3\to 3$ T-matrix, which brings the 
 final state interaction between the three mesons to the 
 $D^\pm\to K^\mp\pi^\pm\pi^\pm$ decay  
is the resonant $S$-wave $I_{K\pi}=1/2$ and the nonresonant one for $I_{K\pi}=3/2$. The
interaction of the identical pions is neglected. In the resonant $K\pi$ channel,
besides the $K^*_0(1430)$ used in parametrization of the LASS data \cite{AstNPB88} given
in Ref.~\cite{E791}, we introduce the resonances $K_0^*(1630)$ (in
PDG there is no assignment of spin to $K(1630)$) and $K_0^*(1950)$
in the $S$-wave $K\pi$ scattering amplitude.

The suggestion to include the higher radial excitations of $K^*_0$
comes in analogy of a recent proposal to interpret the  scalar meson family
($f_0$) as radial excitations of the $\sigma$ meson \cite{dePaulaPLB10} and \cite{MasPRD12}. 
Then, we  make a fit of the LASS $S$-wave phase-shift data in the isospin channel $1/2$
with 
the resonances $K_0^*(1430)$, $K_0^*(1630)$ and $K_0^*(1950)$, which are suggested to be
radial excitations of $K_0^*(800)$. In our anzats the $K\pi$ isospin $1/2$ $S$-wave
S-matrix is given by:
\begin{eqnarray}
S^{\mbox{\tiny{1/2}}}_{K\pi}\left(M^2_{K\pi}\right)= \frac{k\cot\delta+i \,k}{k\cot\delta-i
\,k}\, \prod_{r=1}^3 \frac{M^2_r-M^2_{K\pi}+i z_r \bar\Gamma_r}
 {M^2_r-M^2_{K\pi}-i z_r\Gamma_r}, 
\label{skpi12}
\end{eqnarray}
where $ z_r=k\,M^2_r/(k_r\,M_{K\pi})$, $k\cot\delta=\frac{1}{a}+\frac12 r_0\, k^2$,  and effective range 
parameters  $ a=1.6$~GeV$^{-1}$, $r_0=3.32$~GeV$^{-1}$.
 The relative momentum of the K$\pi$ pair is $k$ and $k_r$ is the momentum at the resonance. 
 \begin{table}[!htb]
\caption{$S$-wave resonance parameters for the $K\pi$ isospin $1/2$ S-matrix. Particle
Data Group (PDG) \cite{PDG}.}
\centering
\label{tab:1}       
\begin{tabular}{llllll}
\hline\noalign{\smallskip}
$\frac12(0^+)$ & $M_r$ (GeV) & PDG (GeV) & $\Gamma_r$ (GeV) & $\bar\Gamma_r$ (GeV) & PDG
(GeV) \\[3pt]
\tableheadseprule\noalign{\smallskip}
$K_0^*(1430)$    &   1.48 &$1.425\pm .050$  & 0.25  &  0.25 & $0.27\pm.08$    \\
$K_0^*(1630)$    &   1.67  & $1.629 \pm .007$ & 0.1 &  0.1 & $< .025$ \\
$K_0^*(1950)$    &   1.9 &$1.945\pm 0.22$  & 0.2 &  0.14 &$0.201\pm0.086$ \\
\noalign{\smallskip}\hline
\end{tabular}
 \end{table}
 
\begin{figure}[!htb]
\centering
\includegraphics[scale=0.25]{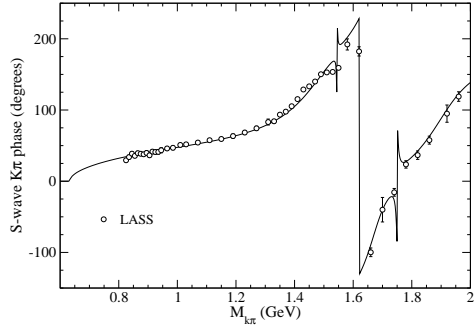}
\caption{$S$-wave $K\pi$ phase as a function of the $K\pi$ mass. Solid
line: phase-shift from the $K\pi$ S-matrix, Eq.~(\ref{skpi12}).
Circles: LASS data \cite{AstNPB88}.} \label{lass}
\end{figure}

 In Fig.~\ref{lass}, we show the results from the three-resonance model, Eq.~(\ref{skpi12}), with parameters given in
 Table \ref{tab:1}.
The $K\pi$ $S$-wave phase-shift is compared to that of LASS. Although the parameters of
the
model are not yet optimized, we are able to reproduce the LASS data reasonably well.  

The non-resonant isospin $3/2$ $S$-wave $K\pi$ S-matrix,  $S^{\mbox{\tiny{3/2}}}_{K\pi}=
\frac{k\cot\delta+i \,k}{k\cot\delta-i\,k}$, 
 is well parametrized 
by the effective-range expansion $k\cot\delta=\frac{1}{a}+\frac12 r_0\, k^2$,  with $a=-1.00$~GeV$^{-1}$
and $r_0=-1.76$~GeV$^{-1}$  according to Ref. \cite{EstNPB78}. 
The $S$-wave K$\pi$ amplitude in the isospin channel $I_{K\pi}$ is written as 
\begin{eqnarray}
\tau_{I_{K\pi}}\left(M_{K\pi}^2\right)=4\pi\,\frac{M_{K\pi}}{k}
\left(S^{\tiny{I_{K\pi}}}_{K\pi}-1\right), \label{tau32}
\end{eqnarray}
which is the input of the three-body rescattering model for  $I_{K\pi}=1/2$ and $3/2$.

\section{Three-Body Rescattering Model}
\label{tbrm}

\begin{figure}[!htb]
\centering
\includegraphics[scale=0.6]{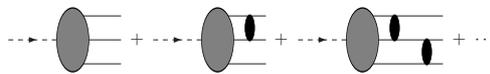}
\vspace{-0.5cm}
\caption{Diagrammatic representation of the heavy meson decay
process into $K\pi\pi$, starting from the partonic amplitude (gray)
and adding the hadronic multiple scattering in the ladder
approximation. The input $K\pi$ scattering amplitude (black) is
required fully off-mass-shell.}
\label{ladder}
\end{figure}

The partonic amplitude for the decay of the $D$ meson into the
$K\pi\pi$ channel with off-shell momenta $q^\mu_i$, and masses $m_i$
$(i=\pi,K,\pi^\prime)$ is expressed by the function
$D(q_\pi,q_{\pi^\prime})$.  It corresponds to a smooth
background given by the direct partonic decay amplitude and is
represented by the gray blob with three legs at leftmost corner of
Fig.~\ref{ladder}. This amplitude should be convoluted with
the $3\to 3$ transition matrix, which take into account the
three-meson interacting final state, as shown in Fig.~\ref{ladder} in a form
of connected ladder series, where the $2\to 2$ scattering process is
summed up in the $K\pi$ transition matrix. In addition, the interaction between the two positively 
charged pions is disregarded.
\vspace{-.5cm}

The full decay amplitude represented diagrammatically in Fig.~\ref{ladder} is given by
\begin{multline}
{\cal A}_0(k_{\pi},k_{\pi^\prime})= a_{0}(s_{A})e^{i\phi_{0}(s_{A})} +
a_{0}(s_{B})e^{i\phi_{0}(s_{B})} = D(k_{\pi},k_{\pi^{\prime}}) + a(m^{2}_{12}) +
a(m^{2}_{23}) \\ 
= D(k_{\pi},k_{\pi^\prime}) +\int \frac{d^4q_\pi d^4q_{\pi^\prime}}{(2\pi )^8}T_{\mbox{\tiny{3,3}}}
(k_{\pi},
k_{\pi^\prime};q_{\pi},q_{\pi^\prime})S_\pi(q_\pi)
S_{\pi}(q_{\pi^\prime})S_K(K-q_{\pi^\prime}-q_{\pi})
D(q_\pi,q_{\pi^\prime}) \, ,\label{D}
\end{multline}
where the momentum of the pions from the decay of the $D$ are
$k_\pi$ and $k_{\pi^\prime}$. The matrix element of the $3\to 3$
transition matrix is $T(k_{\pi},
k_{\pi^\prime};q_{\pi},q_{\pi^\prime})$. The mesonic Feynman
propagators are  $S_i(q_i)=i(q_i^2-m_i^2+i
\varepsilon)^{-1}$, in the approximation where self-energies are
disregarded. The T-matrix operator acts on the isospin space of the
$K\pi\pi$ system, while $D(k_{\pi},k_{\pi^\prime})$ and ${\mathcal
D}(k_{\pi},k_{\pi^\prime})$ are states in the corresponding isospin
space. Our model introduces a two-body transition matrix with matrix elements dependent only on the
Mandelstam $s$-variable, which gives a considerable simplification in the Faddeev decomposition 
of the $3\to3$ off-shell transition amplitude.  Once this assumption is done, one easily
recognizes the factorization
 \begin{equation}
 a(M^{2}_{K\pi}) =\tau\left((K-q_{\pi'})^2\right)\xi(q_{\pi'}) \, , \quad  M^{2}_{K\pi}=(K-q)^2,
\end{equation} 
by inspecting the scattering series that ends with a two-body transition matrix (see Fig. \ref{ladder}), which only 
carries  the dependence on $s_{K\pi}=M^2_{K\pi}$ , and  remains only the dependence on the on-mass-shell momentum of the
bachelor pion ($\pi'$). We have depicted only the $K\pi$ pair, as we additionally assume that the identical pions interact weakly and
the corresponding transition matrix  disregarded.  The total momentum $K$ of the decaying $D-$meson is shared by the $K\pi\pi$ system.

The light-front projection of the covariant equations for the Faddeev components of the decay amplitudes associated 
with the spectator functions decomposed in isospin states for angular momentum zero are written as:
\begin{multline}
\xi^{I_T^z}_{I_T,I_{K\pi}}(y,k_\bot)=\left<I_T,I_{K\pi},I_T^z|D\right>
\xi_0(y,k_\bot)  +\\
+\frac{i}{2 }\sum_{I_{K\pi^\prime}}R^{I_T^z}_{I_T,I_{K\pi},I_{K\pi^\prime}}
\int_0^{1-y}\frac{dx}{x(1-y-x)}\int_0^\infty \frac{dq_\bot  }{(2\pi)^3} 
 K_{I_{K\pi^\prime}}(y,k_\bot;x,q_\bot)\,\xi^{I_T^z}_{I_T,I_{K\pi^\prime}}(x,q_\bot), \label{qsi}
\end{multline}
where the kinematics was chosen such that the decay plane is transverse to $z$-direction, and rotational invariance is preserved after the light-front projection.
The spectator function depends on the total isospin and on the $K\pi$ subsystem isospin quantum numbers, $I_T$ and $I_{K\pi}\, (I_{K\pi'})$  , respectively.
The free squared mass of the $K\pi\pi$  system is
\begin{equation}
M_{0,K\pi\pi}^2(x,q_\bot,y,k_\bot) = \frac{k_\bot^2 + m_\pi^2}{y}
+ \frac{q_\bot^2 + m_\pi^2}{x} 
+ \frac{q_\bot^2 + k_\bot^2 + 2q_\bot k_\bot \cos\theta+ m_K^2}{1-x-y},
\end{equation}
and the squared-mass of the virtual $K\pi$ system is
$M_{K\pi}^2(z,p_\bot) =(1-z)\left(M_D^2 - \frac{p_\bot^2 +
m_\pi^2}{z}\right) - p_\bot^2. $
The isospin recoupling coefficient is
$R^{I_T^z}_{I_T,I_{K\pi},I_{K\pi^\prime}}=\left<I_T,I_{K\pi},I_T^z|I_T,I_{K\pi^\prime},I_T^z\right>$, and the
kernel is given by
\begin{eqnarray} 
K_{I_{K\pi^\prime}}(y,k_\bot;x,q_\bot)=\int_0^{2\pi}
d\theta \,\,
{q_\bot\,\tau_{I_{K\pi^\prime}}\left(M_{K\pi^\prime}^2(x,q_\bot)\right)\over
M_D^2-M_{0,K\pi\pi}^2(x,q_\bot,y,k_\bot)+i\varepsilon}.
\end{eqnarray}
The driving term is
\begin{small}
\begin{eqnarray}
\xi_0(y,k_\bot) =\lambda(\mu^2)+
\frac{i}{2 }\int_0^1\frac{dx}{x(1-x)}\int_0^{2\pi} d\theta
\int_0^\infty \frac{dq_\bot q_\bot}{(2\pi)^3} 
\left[\frac{1}{M_{K\pi}^2(y,k_\bot)-M_{0,K\pi}
^2(x,q_\bot) + i\varepsilon} -\frac{1}{\mu^2-M_{ 0,K\pi}^2(x,q_\bot)}\right] , 
\nonumber
\end{eqnarray}
\end{small}
\hspace{-0.2cm}
where the loop-divergence is regulated by a subtraction and the renormalized value of the amplitude at the subtraction scale is
$\lambda(0)=0.12+0.06i$ for $\mu^2=0$,  which matches the driving term obtained with
the $\chi$-model of the $D\to K\pi\pi$ decay
computed in Ref.~\cite{MagPRD11}. In our actual computations we will allow the subtraction scale to move, by   
keeping $\lambda(\mu^2)$ fixed to $\lambda(0)$, while moving~$\mu^2$.  The detailed derivation of the light-front projected Faddeev equations, 
using the quasi-potential method (see e.g.~\cite{FreFBS11}) will be presented elsewhere.

\section{Results for $D^\pm\to K^\mp \pi^\pm \pi^\pm $ Model Against  Experimental  Analysis}

The previous calculation of the $\chi-$model for the $D$ decay with only isospin $1/2$
$K\pi$ interaction 
calculated the spectator function  in perturbation theory. 
The driving term was responsible for about 70-80\% of the decay amplitude while the next iteration gave the rest. 
That calculation stopped in the next to leading order term. We have checked that indeed the third order term 
contribution to the spectator function are of the order of few percents of 
the total and it explicit from is obtained by iterating twice Eqs.~(\ref{qsi}). 
To obtain the bachelor amplitude a small and
finite imaginary term ($\epsilon=0.2$ GeV) was introduced in the three-meson propagator,
it also represents absorption to other decay channels, which is beyond the model. An
arbitrary subtraction point was chosen for the driving term, with values ranging from $-1\, \text{GeV}^2\lesssim \mu^2\lesssim (m_\pi+m_K)^2$ .
Given that, we solve perturbatively up to the second order the spectator function and computed the
decay amplitude by fully considering the   $K\pi$ interaction in both isospin channels, and written explicitly as:
\begin{eqnarray}
A_0(M^2_{K\pi})=a_0(M^2_{K\pi})e^{i\Phi_0(M^2_{K\pi})}= \sum_{I_T,I_{K\pi}} C_{I_T,I_{K\pi}}\left[\frac{A_{I_T,I_{K\pi}}}{2}
+\tau_{I_{K\pi} }(M^2_{K\pi})\xi^{\mbox{\tiny{3/2}}}
_{I_T,I_{K\pi}}(k_{\pi^\prime}) \right] \ , \label{amplitude}
\end{eqnarray} 
where 
$C_{I_T,I_{K\pi}}=\left<K^-\pi^+\pi^+\right|\left.I_T,I_{K\pi},I_T^z=3/2\right>,$
and $A_{I_T,I_{K\pi}}=\frac{1}{2}\left<I_T,I_{K\pi},I_T^z=3/2\right|\left.D\right>$. Due to the symmetry of the
amplitude against the exchange of the identical pions one are left with
$A_{3/2,3/2}=\sqrt{5}$, $A_{3/2,1/2}=\sqrt{\frac{5}{54 }}(W_1-W_2),$ and
$A_{5/2,3/2}=\frac{W_3}{\sqrt{5}}$. 
In the particular case of $\left|D\right>=\left|K^-\pi^+\pi^+\right>$ one has that 
$W_1=W_2=W_3=1$.  
 
\begin{figure}[!htb]
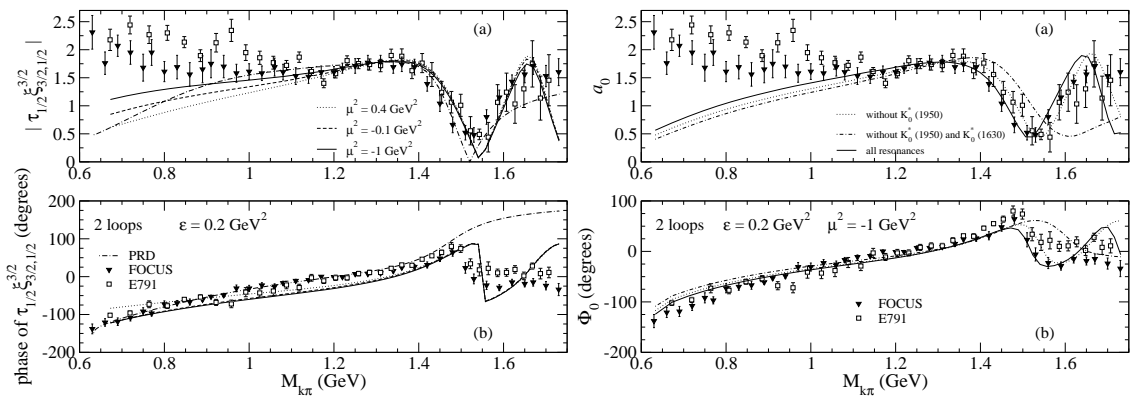

\centering
\includegraphics[scale=0.3]{tauqsi-singlechannel.eps}
\includegraphics[scale=0.3]{fittings-nores.eps}
\caption{{\it Left panel:} Single-channel calculation up to 2-loops for $I_T=3/2$  considering the resonances $K^*_0(1430),\,K^*_0(1630),\,K^*_0(1950)$.
(a) Modulus and (b) phase of
$\tau_{\mbox{\tiny{1/2}}}\xi^{\mbox{\tiny{3/2}}}_{\mbox{\tiny{3/2,1/2}}}$. Values for
$\mu^2$ in GeV$^2$: $0.4$ (dotted line), $-0.1$ (dashed-line), and $-1$ (solid line).  
{\it Right panel:} Coupled-channel calculation up to 2-loops considering $I_T=3/2$ and
5/2.  $K\pi$ interaction isospin $3/2$ and $1/2$  states with  
resonances $K^*_0(1430),\,K^*_0(1630),\,K^*_0(1950)$. The fitted parameters are
 $W_1=1$, $W_2=2$ and $W_3=0.2$. (a) Modulus and (b) phase of the $D\to K\pi\pi$ amplitude for two cases: i)
without $K^*_0(1950)$ and ii) without $K^*_0(1950)$ and $K^*_0(1630)$.
The experimental data come from the phase-shift analysis 
of E791 \cite{E791} (empty box) and FOCUS collaboration \cite{FOCUS2} (inverted full triangle).
(For
reference, the dotted-dashed line gives the previous covariant calculation up to two-loops
from \mbox{Magalh\~aes, \it{et. al.}} of Ref.~\cite{MagPRD11}).}
\label{figmodulusphase}
\end{figure}

The fit found in Ref.~\cite{MagPRD11} below $K^*_0(1430)$ with $K\pi$ interaction in $1/2$
isospin state only, suggested that the partonic
amplitude has little overlap with the $K^\mp\pi^\pm\pi^\pm$ final state channel, i. e.,
the first term in left-hand-side of Eq.~(\ref{amplitude}) should vanishes. 
Here, we also present results computed only by considering $A_0(M^2_{K\pi})\approx
\tau_{\mbox{\tiny{1/2}}}(M^2_{K\pi})\xi^{\mbox{\tiny{3/2}}}_{\mbox{\tiny{3/2,1/2}}}
(k_{\pi^\prime})$. The modulus and phase of this amplitude are shown in the left panel of
Fig.~\ref{figmodulusphase} and compared to the experimental analysis from E791 \cite{E791}
and FOCUS collaboration \cite{FOCUS2}. As in the previous work~\cite{MagPRD11}, a
reasonable fit to the experimental data below $K^*_0(1430)$ is found. However, note that a
structure in the phase is seen in the model which incorporates $K^*_0(1630)$ and
$K^*_0(1950)$, as also verified in the LASS data. A better fit of the LASS data above
$K^*_0(1430)$ seems necessary to find a better description of the valley in the modulus
and the structure of the phase. The conclusion is somewhat independent on the subtraction
point, at least for range of values in the figure.

In the right panel of Fig.~\ref{figmodulusphase}, we present results for the coupled
channel model with $K\pi$ interaction in isospin $1/2$ and $3/2$ states using Eqs.
(\ref{qsi}) up to two-loops. We considered some variation of $\mu^2=0.4,-0.1$ and
$1$~GeV$^2$ in the driven term and fixed $\epsilon=0.2$ GeV$^2$. A reasonable fit of the
experimental phase and modulus is given by $\mu^2=-1$~GeV$^2$ and $\mu^2=-0.1$~GeV$^2$
with $W_1=1$, $W_2=2$ and $W_3=0.2$ in the decay amplitude (\ref{amplitude}) and spectator
functions coupled equations (\ref{qsi}). At low $M_{K\pi}$ below $1$~GeV, the model tends
to underestimate the modulus, where the different analyses of E791 and FOCUS present a
large dispersion.  The characteristics valley and the follow-up height is somewhat
described by the model, with exception of the region close to the boundary of the decay
phase-space, where the data seems to indicates an increase of the amplitude and the model
presents a noticeable decrease. Notice also that the effect of the resonances in the fit
of the $K\pi$ isospin $1/2$ amplitude to the LASS data, in the last model results, is
similar to the single channel case we have already discussed. The region close to the
valley appearing in the modulus is sensitive mainly to our fit of the LASS data in the
neighborhood of $K^*_0(1630)$, while $K^*_0(1950)$ presents a smaller effect in part due
to the competition with the interaction in the $I_{K\pi}=3/2$ state.

\section{Summary and Outlook}

The final state interaction (FSI) contribution to charged $D$ decay into $K\pi\pi$ is 
formulated within a three-body model with Faddeev-like equations for the components of 
the decay amplitude. The dynamical equations for the bachelor amplitudes were computed
within a light-front framework assuming the dominance of the valence state composed by a  $K\pi\pi$ system.  
The  importance of the $S$-wave $I_{K\pi}=3/2$ and $1/2$ interactions ($I_{K\pi}=1/2$ is
dominant) in $D^\pm\to K^\mp \pi^\pm\pi^\pm$  decay is studied against the experimental
phase-shift analysis \cite{E791,FOCUS2}. The convergence of the rescattering series was
checked computing terms up to the third perturbative order, which can be neglected in the
region of parameters used in the fits. The role of the resonances  above $K^*_0(1430)$ and
 the contribution of the $K\pi$ isospin $3/2$ channel to charged three-body $D$ decays
were studied and we found that the not established resonance $K^*_0 (1630)$  with quantum
numbers $\frac{1}{2}(0^+)$ and width $\Gamma \lesssim 100 $ MeV is necessary to improve
the fit of the phase and magnitude  of the charged $D\to K\pi\pi$ unsymmetrized decay
amplitude above $K^*_0 (1430)$.  In addition, we confirmed the importance of the isospin
$3/2$ contribution to this decay \cite{MagPOS12}.

One interesting application of the formalism is the study of the CP violation in
three-body charged $B$ decays, where recent experiments \cite{LHCB1} showed a sizable
effect, interpreted in Ref.~\cite{BedArX13} as resulting from the interference between the
weak phase and the strong phase from the final state interaction. 

\begin{acknowledgements}
We thank the Brazilian funding agencies FAPESP
(Funda\c{c}\~{a}o de Amparo a Pesquisa do Estado de S\~{a}o Paulo),
  CNPq (Conselho Nacional de Pesquisa e
Desenvolvimento of Brazil). 
\end{acknowledgements}



\end{document}